\pdfoutput=1
\documentclass{sig-alternate}

\usepackage{blindtext, graphicx}
\usepackage{amsmath}
\usepackage[linesnumbered,boxruled]{algorithm2e}
\usepackage{comment}
\usepackage{verbatim}
\usepackage{subcaption}
\usepackage{url}
\newcommand{\discsetseqdec}{D{\scriptsize ISC}-S{\scriptsize ET}-S{\scriptsize EQ}-D{\scriptsize EC}\xspace}

\newcommand{\tuple}[1]{\ensuremath{\left\langle#1\right\rangle}\xspace}
\newcommand{\completiondec}{C{\scriptsize OMPLETION}-D{\scriptsize
EC}\xspace}
\newcommand{\cnfsat}{C{\scriptsize NF}-S{\scriptsize AT}\xspace}
\newcommand{\cnp}{\ensuremath{\mathbf{NP}}\xspace}
\newcommand{\cpspace}{\ensuremath{\mathbf{PSPACE}}\xspace}

\newtheorem{definition}{Definition}

\newtheorem{theorem}{Theorem}
\newtheorem{proposition}{Proposition}

\begin{document}
%
\title{Reverse Engineering Camouflaged Sequential Circuits Without Scan Access}



%
\numberofauthors{3}
\author{
%
%
\alignauthor
Mohamed El Massad\\
       \affaddr{New York University Tandon School of Engineering}\\
       \affaddr{6 MetroTech Center}\\
       \affaddr{Brooklyn, New York 11201}\\
       \email{me1361@nyu.edu}
\alignauthor
Siddharth Garg\\
       \affaddr{New York University Tandon School of Engineering}\\
       \affaddr{6 MetroTech Center}\\
       \affaddr{Brooklyn, New York 11201}\\
       \email{sg175@nyu.edu}
\alignauthor Mahesh Tripunitara
       \affaddr{Electrical and Computer Engineering Department}\\
       \affaddr{University of Waterloo}\\
       \affaddr{Waterloo, Ontario N2L 3G1}\\
       \email{tripunit@uwaterloo.ca}
}


\maketitle

\begin{abstract}
Integrated circuit (IC) camouflaging is 
a promising technique to 
protect the design of a chip from 
reverse engineering. 
However, 
recent work has shown that even camouflaged ICs 
can be reverse engineered 
from the observed input/output 
behaviour of a chip using SAT solvers. 
However, these so-called 
SAT attacks have so far targeted only camouflaged 
combinational circuits. For 
camouflaged sequential circuits, the SAT attack requires
that 
the internal state of the circuit is controllable and observable via the scan chain.
It has been implicitly
assumed 
that 
restricting scan chain access increases the security of 
camouflaged ICs from reverse engineering attacks.
In this paper, 
we develop a new attack methodology 
to decamouflage sequential circuits without scan access.  
Our attack uses a model checker (a more powerful reasoning tool than a SAT solver)
to find 
a discriminating set of input \emph{sequences}, i.e., one 
that is sufficient to determine the functionality of camouflaged gates. 
We propose several refinements, including the use of a bounded model checker, 
and sufficient conditions for determining when a set of input sequences is discriminating 
to improve
the run-time and scalabilty of our attack. 
Our attack is able to decamouflage a large sequential benchmark 
circuit that implements a subset of the VIPER processor. 
\end{abstract}

\section{Introduction}
Vendors that provide commercial IC reverse engineering services are an increasing threat to the confidentiality of 
IC designs. 
Using chemical etching and high-resolution microscopy,
vendors of reverse engineering services  
have reconstructed gate-level netlists of complex nanometer scale ICs~\cite{chipworks}, thus compromising 
the IC designer's intellectual property (IP). 
IP theft of 
this nature can negatively impact an IC designer's 
revenue and competitive advantage.

IC camouflaging 
is a
promising 
technique to protect the designer's IP against 
reverse engineering attacks. 
IC camouflaging 
works by augmenting a traditional CMOS technology library with 
so-called \emph{camouflaged} 
standard cells. 
A camouflaged standard cell can implement one of many Boolean logic functions, 
even though its layout looks the same 
to a reverse engineer regardless of its functionality. 
Several different techniques 
have been proposed in literature to implement
camouflaged standard cells. These include the use of dummy contacts~\cite{syphermedia} and 
threshold-voltage dependent camouflaging~\cite{larson1992convertible,walden1993dynamic,collantes2016threshold}. 
Given the economic and strategic value of IC camouflaging, 
there has been considerable research on determining which gates in an IC to camouflage so as to maximize 
security~\cite{RSS+13,yasin2016camoperturb,li2016provably}.

Recently,
El Massad et al.~\cite{el2015integrated} 
demonstrated that all existing camouflaging schemes can be broken using the so-called ``SAT attack." (A similar technique 
to defeat logic encryption was concurrently proposed in ~\cite{subramanyan2015evaluating}.)
The SAT attack assumes that the attacker has access to two functioning copies of the IC. One is reverse-engineered 
to reconstruct the IC's netlist barring, of course, 
the functionality of each camouflaged standard cell. The other copy 
is used to observe the IC's input/output (I/O) behavior.

\begin{figure*}
\centering
  \includegraphics[width=1.05\textwidth]{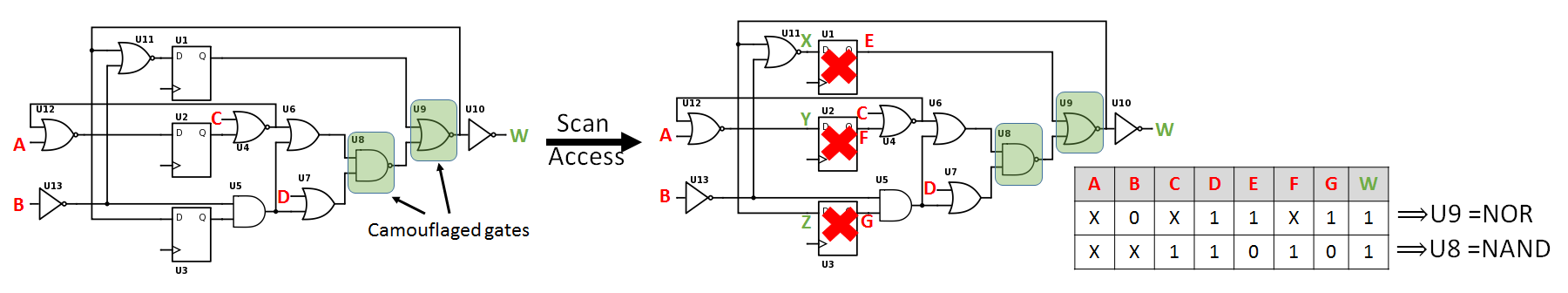}
\caption{ISCAS s27 sequential benchmark circuit with two camouflaged gates, $G_1$ and $G_2$. 
Also shown are the two states in the FSMs corresponding to the correct decamouflaging solution 
and incorrect solutions, respectively.}
\label{fig:scanaccess}
\end{figure*}

The goal of the SAT attack is to find a 
set of inputs 
that are \emph{sufficient} 
to deduce the functionality of the camouflaged standard cells. 
The attack iteratively determines new inputs 
that 
prune the 
attacker's search space. The attack terminates 
when the camouflaged standard cells 
have only one unique assignment (or multiple functionally 
equivalent assignments). 
New input patterns and the final completion (a completion is an assignment of identities to camouflaged gates) 
are determined using a SAT solver. 
El Massad et al. demonstrated empirically that only a small number of inputs are required 
to exactly decamouflage even 
large benchmark netlists, and that attackers can do so in the order of minutes. 
This work has resulted in renewed focus on 
stronger camouflaging schemes that are secure against SAT attacks~\cite{yasin2016camoperturb,li2016provably}.  


However, one criticism of the SAT attack and its subsequent 
enhancements~\cite{liu2016oracle} is that these attacks
have all
focused on decamouflaging 
combinational netlists, i.e., 
netlists without internal state. 
Real world ICs, on the other hand, are typically
\emph{sequential}, i.e., they 
implement finite state machines (FSMs) with internal state 
stored in flip-flops.  
The SAT attack implicitly
assumes that the IC's internal state
can be fully controlled and 
observed via scan chains, thus 
reducing the problem to that of
reverse engineering a combinational 
netlist. 
A designer concerned about IP theft, however, 
can easily block user-mode 
access to the scan chain using a 
secure scan interface~\cite{lee2006low}.
Thus, although the attacker still has access to the IC's 
primary I/Os, 
they \emph{cannot} control or 
observe the
state of internal flip-flops during IC operation. 
As we illustrate below, 
\textbf{the SAT attack does not work for ICs with internal 
state that cannot be accessed via scan chains}. 
This is a major practical limitation of the SAT attack. 

\begin{figure}
\centering
  \includegraphics[width=0.9\columnwidth]{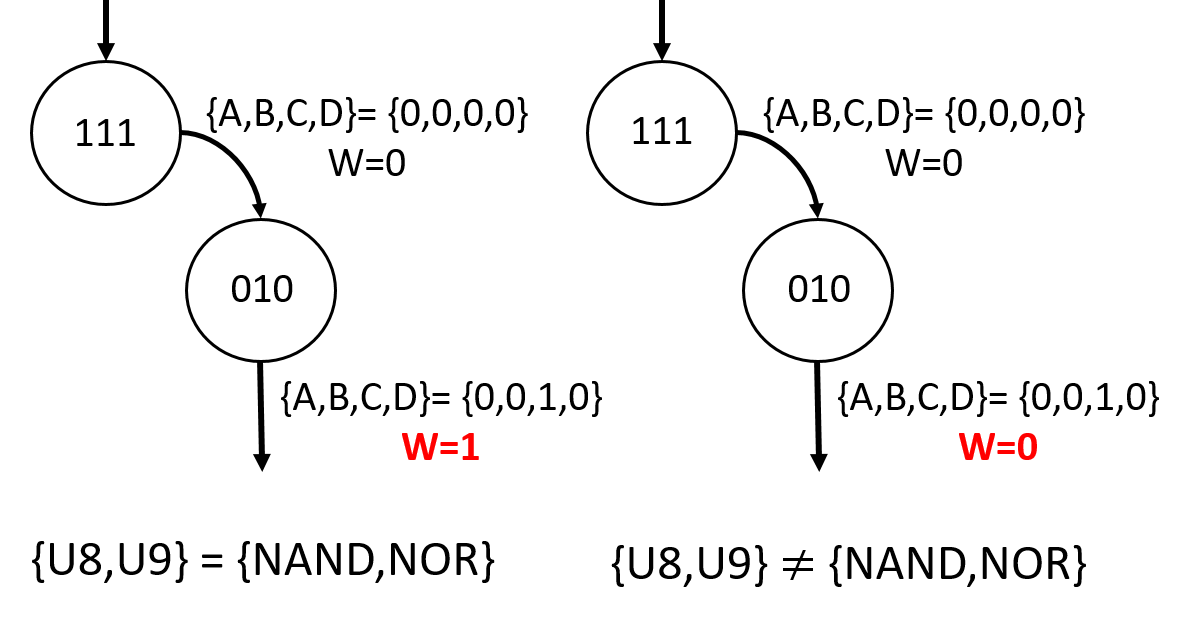}
\caption{ISCAS s27 sequential benchmark circuit with two camouflaged gates, $G_1$ and $G_2$. 
Also shown are the two states in the FSMs corresponding to the correct decamouflaging solution 
and incorrect solutions, respectively.}
\label{fig:noscanaccess}
\end{figure}

\noindent \textbf{Motivational Example}
Consider the netlist shown in Figure~\ref{fig:scanaccess}. 
The netlist corresponds to the s27 circuit from the ISCAS'89 sequential 
benchmark suite\cite{iscas89}. 
Two gates, NAND gate U8 and a NOR gate U9 
have each been implemented using a camouflaged standard cell.
We assume that the camouflaged standard cell can implement 
\emph{either} a NAND or a NOR gate.
The primary output W of the circuit depends on both the  
primary inputs ($A$, $B$, $C$, and $D$) 
and the state of the flip-flops. We will assume, without any loss of generality, that all flip-flops
are initially set using a global (re)set signal.


As shown in Figure~\ref{fig:scanaccess}, an attacker with full scan access
can set the output of each flip-flop to any desired value, and can thus treat flip-flop outputs
as new primary inputs (inputs $E$, $F$ and $G$). Similarly, the attacker can 
scan out the the input of each flip-flop, 
and can thus treat flip-flop inputs as new primary outputs (outputs X, Y and Z). The SAT attack, or in fact even the less powerful
logic testing based attack proposed by Rajendran et al.~\cite{RSS+13}, can be used to decamouflage  the 
resulting combinational netlist. Figure~\ref{fig:scanaccess} shows two 
inputs that are sufficient to decamouflage gates U8 and U9; the first input reveals the identity of U8
while the second reveals the identity of U9.

Now consider an attacker \emph{without} scan access. The attacker can no longer control 
flip-flop outputs
or observe flip-flip inputs. To apply the SAT attack, 
an attacker can treat the sequential 
circuit as a single-stage combinational circuit by repeatedly resetting the flip-flops, applying primary inputs $A$, $B$, $C$ and $D$,  
and observing the primary output $W$. 
Unfortunately, this strategy does not work. As shown in Figure~\ref{fig:noscanaccess}, primary output $W$ equals $0$ 
regardless of the identity of 
camouflaged gates U8 and U9.

Yet, as shown in Figure~\ref{fig:noscanaccess}, applying a 
\emph{sequence} of two inputs 
recovers the correct identities of the two camouflaged gates. That is, the output of the FSM after the second 
input is applied is $1$ if and only if gate U8 is a NAND and gate U9 is a NOR.
In general, we note that an attacker might 
require not only one but 
multiple input sequences to reverse engineer camouflaged 
sequential circuits without scan access. 
Finding a set of input sequences that is sufficient to decamouflage the netlist is the goal 
of our attack.




In general, removing scan access from a camouflaged sequential circuit makes 
the reverse engineering problem more challenging for several reasons. 
For one, an attacker 
with scan access can arbitrarily set 
the state of the sequential circuit to states that help 
discriminate the identities of camouflaged gates (as in the example above). 
On the other hand,   
an attacker without scan access must apply
a sequence of inputs that lead the sequential circuit to the desired state. 
However, the input sequence itself depends on the 
identities of camouflaged gates. Second, an attacker without scan access 
only observes the primary outputs and
must \emph{infer} 
the next state outputs 
computationally. Given these challenges, a natural question that arises
is the following: 
{does restricting scan chain access for camouflaged sequential circuits enhance their security
against reverse engineering attacks?}
Our new attack seeks to answer this question both foundationally and empirically.

\noindent \textbf{Our Contributions}
In this paper, we make the following novel contributions.
\begin{itemize}

\item We introduce the first attack methodology 
to decamouflage \emph{sequential}  
circuits without access to internal state of the flip-flops.  
Our attack searches iteratively for \emph{input sequences}; each  
new input sequence eliminates one or more remaining decamouflaging solutions 
till only correct completions/solutions are remaining.

\item We characterize the computational complexity of two important sub-problems in our 
attack procedure: is a given set of input sequences sufficient to decamouflage 
the netlist (\discsetseqdec), and finding a completion that is consistent with a set of input sequences (\completiondec). 
We show that the former problem 
is in \textbf{PSPACE}, while the latter is \textbf{NP}. 
Consequently, our attack uses a 
\emph{model checker}  
to find new input sequences, to decide when to 
terminate and to identify correct completions. 

\item We propose a practical attack methodology that utilizes a 
\emph{bounded} model checker and \emph{sufficient} conditions 
for the \discsetseqdec problem 
to reduce the run-time of the attack.  

\item Our 
experimental results 
speak to the strength of our attack; we are able to decamouflage a sequential benchmark that represents 
a part of the \emph{VIPER} processor netlist with 
more than 5000 gates in a matter of hours. 
For benchmarks which our attack fails to 
fully decamouflage, we still correctly decamouflage up to 30 out of 32 camouflaged gates.   
\end{itemize}
\section{Related Work}

Techniques for extracting the underlying netlist of integrated circuits via
chemical etching, delayering and
scanning electron microscopy (SEM) are offered by
companies like Chipworks~\cite{chipworks} and Degate~\cite{degate} as part of their commercial reverse-engineering services. 
These companies also develop and offer software tools to aid in the process of circuit extraction. 
Torrence et al. \cite{torrance2009state} provide a detailed overview of the IC reverse engineering process.

Camouflaging technology aims to protect against the misuse of these IC reverse-engineering techniques for piracy and copyright infringement. Several proposals have been made, both in academia and industry, for implementing camouflaged cells for use in ASIC processes. These include dummy-contact-based camouflaged cells~\cite{syphermedia,rajendran2012security,li2016provably} as well as threshold-voltage dependent gates~\cite{collantes2016threshold,li2016provably}. 


Because camouflaged standard cells incur area, delay and power overheads, recent research has focused on 
determining which gates to camouflage to maximize security. 
However, all of the work has considered camouflaging only combinational circuits, or equivalently, assumed 
sequential circuits in which the attacker has scan access.
Rajendran et al.~\cite{RSS+13} showed that randomly selecting gates to camouflage is vulnerable to 
VLSI testing based attacks, and proposed a new selection scheme that tries to maximize the number of 
non-resolvable gates.
However, this scheme was broken by El Massad~\cite{el2015integrated} and Subramanyan et al.'s SAT attacks.
In response to these attacks, \cite{yasin2016camoperturb} and \cite{li2016provably} concurrently developed 
SAT-attack resilient schemes that try to ensure that discriminating sets are exponentially sized. 
However, as acknowledged by the authors, these schemes also come with a fundamental trade-off: the 
output corruptability (or error rate) of these schemes is low; that is, incorrect completions agree with 
correct completions on almost all inputs.

In this paper, we seek to analyze
the security of 
camouflaging schemes that purport to defend against SAT attacks in a different way, i.e., 
by removing 
access to scan chains (instead of reducing output corruptability/error rate as in the schemes proposed by \cite{yasin2016camoperturb} and \cite{li2016provably}). While our new attack is successful on a range of benchmarks, we also find that 
there are some benchmarks whose security is enhanced by removing scan access. 

\section{Attack Procedure}
In this section, we describe our attack procedure. 
We first precisely describe the attack objective 
and introduce some notation that aids our exposition. 
Then, we define two computational problems that form 
the foundation of our procedure and characterize their 
computational complexity. Finally, we describe how the procedure works, 
and the practical choices we made while designing the attack.

\newcommand{\mvt}[1]{\textit{\{mvt: #1\}}}

\subsection{Problem Formulation}
As noted before, we assume that the 
attacker obtains two copies of the IC. 
The attacker use the first copy as a black box and exercises it with inputs.
Let $\mathcal{C}$ represent the black-box IC. 
The attacker using chemical etching and imaging to extract the netlist 
of the second IC --- let $C$ refer to the extracted netlist. A subset of gates in 
$C$ are camouflaged. 

Let $m$ be the number of primary inputs, $n$ be the number of primary outputs, $k$ be the number of camouflaged gates and 
$l$ be the number of flip-flops (bits of internal state) in $\mathcal{C}$ (and $C$). For instance, in 
Figure~\ref{fig:scanaccess}, $m=4$, $n=1$,  $k=2$ and $l=3$. 
We assume, without loss of generality, that 
each camouflaged gate in the IC implements one of $t$ Boolean functions. 
In the example in 
Figure~\ref{fig:scanaccess}, for instance, $t=2$ because each camouflaged gate is either a NAND or a NOR. 
A \emph{completion} $X: \{1,2,\ldots,k\} \rightarrow \{1,2,\ldots,t\}$ assigns a Boolean function to each camouflaged gate 
in $C$. Given a completion $X$ of $C$, we denote the completed 
circuit by $C_X$.


Now consider a sequence of inputs 
$I=(i_0,i_1,\ldots,i_{p-1})$ 
of length $p$ applied to $\mathcal{C}$ starting 
from the initial reset state, $s_0$. Here, $i_0$ is the input applied 
in the first time step, $i_1$ is the input applied in the second time step, and so on.  
Let $\mathcal{C}(I)=(\mathcal{o}_0,\mathcal{o}_1,\ldots,\mathcal{o}_p)$ 
denote the sequence of outputs that $\mathcal{C}$ produces for input sequence $I$. 

Similarly, 
for a completion $X$ of $C$, let $C_X(I)=(o_0,o_1,\ldots,o_p)$ denote the sequence of outputs produced by circuit $C_X$ for sequence $I$, 
assuming as above that inputs are applied starting from the initial reset state, $s_0$

Let  $\mathcal{I}$ denote the set of all input sequences of length $2^l$, which we refer to as the 
\emph{universal set of input sequences}.
Given $\mathcal{C}$ and $C$, the goal of our attack is to find a completion $X^*$ such that
\begin{equation}
 \forall I \in \mathcal{I}, \quad C_{X^*}(I) = \mathcal{C}(I),
 \label{eq:goal}
\end{equation}
that is, we seek an assignment $X^{*}$ of Boolean functionalities to camouflaged gates 
such that the outputs of the completed netlist, $C_{X^{*}}$ agree with outputs of the black-box 
circuit $\mathcal{C}$ on \emph{all} input sequences of length $2^{l}$. 
Equivalently, we seek an $X^{*}$ such that $C_{X^{*}}$ is sequentially equivalent to $\mathcal{C}$.
We call such a completion a \emph{correct completion}. 
We note that a correct completion is not necessarily unique. 
However, as we find any correct completion, 
it means we have successfully reverse-engineered (the Boolean functionality of) $\mathcal{C}$.

\subsection{Foundations of Our Attack}
In this section, we define two decision problems that form the foundation of our attack procedure. 
We start by introducing the notion of a \emph{discriminating set of input sequences}, which generalizes 
the notion of a discriminating set of inputs that was introduced by El Massad et al.~\cite{el2015integrated} in the context 
of decamouflaging combinational circuits.

\begin{definition}
A set of input sequences $\mathbf{I} = \{I_0, I_1,\ldots, I_n\}$ is called \emph{discriminating} if every completion $X$ that satisfies

\begin{quote}
    $C_{X}(I_j) = \mathcal{C}(I_j), \quad \forall j \in \{1,\ldots,n\}$
\end{quote}

is a correct completion.
\end{definition}

We now articulate the problem of deciding whether a set of input sequences 
is discriminating for a sequential circuit, and identify the 
computational complexity class to which the problem belongs.

\begin{definition}\label{def:discsetdec}
We define \discsetseqdec to be the following decision problem. Given the
following three inputs:
(i) a camouflaged circuit $C$, (ii) $\mathbf{I}$, a set of input sequences,
and (iii) the set of outputs obtained from applying input sequences
in $\mathbf{I}$ to the black-box circuit, each time starting from state $s_0$, i.e.,
$\mathcal{C}(\mathbf{I}) = \left\{\mathcal{C}(I_1), \ldots, \mathcal{C}(I_n)\right\}$,
where $\mathbf{I} = \left\{I_1, \ldots, I_n\right\}$.
Is $I$ a discriminating set for $C$?
\end{definition}

\begin{theorem}\label{thm:discsetdec}
\discsetseqdec is in \cpspace. 
\end{theorem}
The proof for the above theorem is in the appendix. 
We remark on the significance 
of this result below.


\noindent \textbf{Remark.} 
The fact that \discsetseqdec is in \cpspace 
suggests that a viable strategy for tackling the 
problem is to reduce it to model checking which is known to be complete for \cpspace. This is exactly what we do in our attack procedure, as we describe in Section~\ref{sec:pracattack}.

\noindent \textbf{Remark.} 
The corresponding decision problem for camouflaged combinational circuits, i.e., whether a given set of inputs (note, not input sequences) 
is discriminating, was found to be in the complexity class co-\cnp, which is contained in \cpspace. 

We next define a problem that 
captures the computational task of finding a correct completion given a discriminating set of input sequences 
for a camouflaged circuit. 

\begin{definition}
We define \completiondec to be the following decision problem. Given
the following three inputs:
(i) a camouflaged circuit $C$, (ii) $\mathbf{I}$, a set of input sequences,
and (iii) the outputs obtained from applying inputs
in $\mathbf{I}$ on the black-box circuit, i.e., $\mathcal{C}(\mathbf{I})$.
Does there exist a completion $X$ such that
$\forall I \in \mathbf{I}, C_X(i) = \mathcal{C}(I)$?
\end{definition}

\begin{theorem}\label{thm:completiondec}
\completiondec is in \cnp.
\end{theorem}

\begin{proof}
A certificate for \completiondec is a completion $X$ such that $C_X$ agrees with the black-box circuit on input sequences in $\mathbf{I}$. $X$ is polynomially sized in the input to \completiondec ($X$ can be encoded using $k\log(t)$ bits). Verifying that $C_X$ agrees with the black box on sequences in $\mathbf{I}$ can be done in time $O(|\mathbf{I}||C|)$ which is polynomial in the size of the input.
\end{proof}

We note here that the relevance of 
\completiondec is that it yields a correct completion (as a certificate) 
when the input to the problem is a 
discriminating set of input sequences.  
The theorem implies that one can reduce \completiondec to a problem that is complete for \cnp, in particular, to \cnfsat, and then use an off-the-shelf SAT solver for the resulting SAT instance. 
Our reduction from \discsetseqdec to the model checking problem however facilitates an alternative and more convenient approach for \completiondec, as we describe below 
in Section~\ref{sec:pracattack}.

\subsection{Practical Attack Procedure}
\label{sec:pracattack}

Our attack proceeds iteratively: 
we maintain a set of input 
sequences $\mathbf{I}$ 
that is initially empty. 
In each iteration, we add one 
(or more) new input sequences 
to the set. We stop when we determine that our
set of input sequences is discriminating. 

However, 
a naive implementation of this procedure 
would make a call to an \emph{unbounded}
model checker in each iteration, that is, a model checker 
that searches for input sequences of arbitrary length. 
Unfortunately, calls to an unbounded model checker 
can be time consuming. Instead, we  
add new input sequences 
of \emph{bounded} length (using a bounded model checker), and 
increase the bound only when needed. 
Further, instead of directly calling an unbounded model 
checker to decide if the current set of input sequences 
is discriminating (the termination condition for our 
procedure), we first 
check two simpler 
\emph{sufficient} conditions for termination. 
Our refined attack procedure is described below.

\subsubsection{Finding New Input Sequences}
To find new input sequences to add to our set, 
we construct a solver $M_{BMC}$ for \discsetseqdec by reducing \discsetseqdec to a bounded model checking problem. 
Given $C$, $\mathbf{I}$, and
$\mathcal{C}(\mathbf{I})$ as defined in Definition \ref{def:discsetdec}, and a parameter 
$b$ that specifies the model checking bound, 
$M_{BMC}$ returns \texttt{true} if and only if for any two completions $X_1$ and $X_2$, and every input sequence $I$ of length at most $b$, the following implication is true:

\[ C_{X_1}(\mathbf{I})=C_{X_2}(\mathbf{I})=\mathcal{C}(\mathbf{I})\implies C_{X_1}(I)=C_{X_2}(I).\]

Note that if $M_{BMC}$ returns \texttt{true}, 
it does necessarily mean that the given set 
$\mathbf{I}$ is indeed a discriminating set of input sequences for our camouflaged circuit. 
It means only that the solver is 
unable to find a new input sequence of length at 
most $b$ that helps to eliminate any of the remaining 
completions.

If 
if $M_{BMC}$ returns \texttt{false}
on the other hand, 
it returns, 
two completions $X_1$ and $X_2$ and a 
new input sequence $\tilde{I}$ of length at most 
$b$
such that $C_{X_1}(I)=C_{X_2}(I)=\mathcal{C}(I)$ for all $I \in \mathbf{I}$, but $C_{X_1}(\tilde{I}))\neq C_{X_2}(\tilde{I})$, i.e., $C_{X_1}$ and $C_{X_2}$ agree with the black-box circuit on input sequences in $\mathbf{I}$, but produce different outputs for input sequence $\tilde{I}$.

We call $M_{BMC}$ at every iteration in our algorithm; passing it our camouflaged circuit, our current set of input sequences, and the output of the black-box circuit for each sequence in the set. If $M_{BMC}$ returns with a sequence $\tilde{I}$, we add $\tilde{I}$ to our set of input sequences.

\subsubsection{Termination Criteria}
As we stated previously, our $M_{BMC}$ solver cannot decide whether a set of input sequences $\mathbf{I}$ 
is discriminating. 
For this, we need to call an unbounded model checker.
However, before calling the unbounded model checker, we check for two conditions that are sufficient to show  
that $\mathbf{I}$ is discriminating. The intuition behind performing these 
checks before calling an unbounded model checker is that we expect them to be 
computationally less time consuming.

\begin{enumerate}
    \item \textbf{Unique Completion (UC):} We check to see if there is only one remaining completion that agrees with the black-box circuit on the current set of input sequences. Specifically, we try to find two distinct completions that agree with the black-box circuit on the current set of input sequences, i.e. we try to find two completions $X_1$ and $X_2$ such that 
    
    \[ C_{X_1}(\mathbf{I})=C_{X_2}(\mathbf{I})=\mathcal{C}(\mathbf{I}) \quad \rm{and} \quad X_1 \neq X_2\]
    
    \begin{proposition}
     If no such $X_1$ and $X_2$ exist, then $\mathbf{I}$ is a discriminating set for $C$.
    \end{proposition}
    \begin{proof}
    Follows immediately from the definition of a discriminating set of input sequences.
    \end{proof}
    \item \textbf{Combinational Equivalence (CE)} 
    Next, 
    we check whether all completions that agree with the black-box circuit on the current set of input sequences are \emph{combinationally} equivalent with respect to both output and next state function. 
    That is, if we denote by $\tilde{C}_X(i,s)=(o,s')$ the pair of output $o$ and next state $s'$ of completed circuit $C_X$ when in state $s$, and at the application of input $i$, we ask whether 
    
    \begin{multline}\label{eqn:comb_equiv}
    C_{X_1}(\mathbf{I})=C_{X_2}(\mathbf{I})=\mathcal{C}(\mathbf{I})\implies\\
    \tilde{C}_{X_1}(i,s)=\tilde{C}_{X_2}(i,s) \quad \forall i\in\{0,1\}^m, s\in\{0,1\}^l.
    \end{multline}
    \begin{proposition}
     If Condition (\ref{eqn:comb_equiv}) holds for set $\mathbf{I}$ and camouflaged circuit $C$, then $\mathbf{I}$ is a discriminating set for $C$.
    \end{proposition}
    \begin{proof}
    Since $\tilde{C}_{X_1}(i,s)=\tilde{C}_{X_2}(i,s)$ for all possible inputs $i$ and all possible states $s$, it follows that $C_{X_1}(I)=C_{X_2}(I)$ for every input sequence $I$. Thus, $\mathbf{I}$ is discriminating for $C$.
    \end{proof}
    
    We check Condition (\ref{eqn:comb_equiv}) by calling a solver $CE$ that we construct. We give it as input: (i) the camouflaged circuit $C$, (ii) our current set of input sequences $\mathbf{I}$, and (iii) the outputs $\mathcal{C}(\mathbf{I})$ of the black-box circuit for input sequences in $\mathbf{I}$. If the solver returns \texttt{true}, we terminate. 
    
    \item \textbf{Unbounded Model Check (UMC)} If the preceding two checks fail, we finally call a solver $M_{UMC}$ that we have constructed. The `$U$' is for ``Unbounded.'' We give it as input: (i) our camouflaged circuit $C$, (ii) the current set of input sequences $\mathbf{I}$, and, (iii) the outputs $\mathcal{C}(\mathbf{I})$ of the black-box circuit. The solver $M_{UMC}$, unlike $M_{BMC}$, returns \texttt{true} if and only if $\mathbf{I}$ is a discriminating set for $C$. 
    \end{enumerate}
    
\subsubsection{Finding a Correct Completion}\label{sec:completiondec}
Since the \completiondec problem is in \cnp, it can be reduced to a CNF-SAT instance and 
solved using a SAT solver. However, we note that our
model checking based solver for \discsetseqdec  has the property that 
if $\mathbf{I}$ is a discriminating set for $C$, 
then a correct completion is 
encoded in \emph{every} initial state of the model. 
As such, we do not 
need to call an external SAT solver to find 
a correct completion. We simply ask our model checker to choose any element from the set of initial states. 
We denote this procedure for the \completiondec problem as solver $N$. 
$N$ takes as input the camouflaged circuit $C$, the current set of input sequences $\mathbf{I}$, 
the outputs $\mathcal{C}(\mathbf{I})$ of the black-box circuit, and outputs a correct completion $X^{*}$ if $\mathbf{I}$ is discriminating. 



\subsubsection{Complete Algorithm}
Our complete algorithm is expressed as Algorithm~\ref{alg:reverse_eng}. In the algorithm, we start with an initial bound $b=0$ for our BMC solver, i.e., $M_{BMC}$. At every iteration, if we determine that we need to continue, i.e., all three checks described in the previous section fail, we increase the value of $b$ by a fixed increment $bmc\_inc$, and we continue until at least one of the three checks succeeds. At the end, once we arrive at a discriminating set of input sequences for our camouflaged circuit, we employ the technique described in Section~\ref{sec:completiondec} to find a correct completion for our circuit.

\begin{algorithm}
\DontPrintSemicolon
\SetAlgoNoEnd
$\mathbf{I} \gets \emptyset$, $b \gets 0$\;
\While{\textbf{true}}{
    $b \gets b + \mathit{bmc}\_\mathit{incr}$\;
    $\tuple{X_1, X_2, \tilde{I}} \gets M_{BMC}(C, \mathbf{I}, \mathcal{C}(\mathbf{I}), b)$\;
    \lIf{$\tuple{X_1, X_2, \tilde{I}} \not= \textbf{true}$}{$\mathbf{I} \gets \mathbf{I} \cup \tilde{I}$}
    \lElseIf{$\mathit{UC}(C, \mathbf{I}, \mathcal{C}(\mathbf{I}))$ or $\mathit{CE}(C, \mathbf{I}, \mathcal{C}(\mathbf{I}))$ or
    $M_{\mathit{UMC}}(C, \mathbf{I}, \mathcal{C}(\mathbf{I}))$}{break}
}
    \Return{$N(C, \mathbf{I}, \mathcal{C}(\mathbf{I}))$}

\vspace*{0.0625in}
\caption{Scan-Chain-Free Decamouflaging. 
} \label{alg:reverse_eng}
\end{algorithm}


\subsection{Implementation of Solvers}

We now describe how the solvers referred to in the previous section 
are implemented using a model checker. 
Model checkers take as input a model of an FSM represented as a Kripke structure. 
We begin by describing an FSM model for the camouflaged circuit $C$, 
and the Kripke structure that we use as input
for the model checker. 

Corresponding to each completion $X$ of circuit $C$
is an FSM $(\mathcal{I},\allowbreak \mathcal{O},\allowbreak \mathcal{S},\allowbreak s_0,\allowbreak \sigma_x,\allowbreak \omega_x)$ where:

\begin{itemize}
    \item $\mathcal{I}=\{0,1\}^m$ is the input alphabet of the FSM, the set of all possible inputs,
    \item $\mathcal{O}=\{0,1\}^n$ is the output alphabet of the FSM, the set of all possible outputs,
    \item $\mathcal{S}=\{0,1\}^l$ is the set of states of the FSM, a set of all $l$-bit Boolean vectors,
    \item $s_0 \in S$, the initial state of the black-box circuit,
    \item $\sigma_X$ is the state-transition function, $\sigma_X\!: \mathcal{S}\times\mathcal{I}\rightarrow\mathcal{S}$,
    \item $\omega_X$ is the output function, $\omega\!: \mathcal{S}\times\mathcal{I}\rightarrow\mathcal{O}$.
\end{itemize}

For an input sequence $I=(i_0, i_1,\ldots)$, we can write $C_X(I)$ in terms of $\sigma_X$ and $\omega_X$ as follows: 

\[ C_X(I)=(\omega_X(i_0,s_0),\omega_X(i_1,\sigma_X(s_0,i_0)),\ldots) \]

Our $M_{UMC}$ solver takes an instance $\tuple{C,\mathbf{I},\mathcal{C}(\mathbf{I})}$ of \discsetseqdec and transforms it into a model checking instance as follows. Let $\mathcal{X}=\{0,\ldots,t-1\}^k$ be the set of all possible completions.
We define a set of atomic propositions $AP$ to be a singleton, consisting of an atomic proposition \texttt{equiv}, the semantics of which we clarify below.
We build a Kripke structure $M$ over $AP$ from the tuple $\tuple{C,\mathbf{I},\mathcal{C}(\mathbf{I})}$ that is input to the \discsetseqdec problem. The characteristics of $M$ are as follows: 
\begin{itemize}
    \item The set of initial states of $M$ is $\mathcal{S}\times\mathcal{X}\times\mathcal{S}\times\mathcal{X}\times\mathcal{I}$, i.e., any state in $M$ is a 4-tuple of the form $(s_1,X_1,s_2,X_2,i)$ where $s_1$ and $s_2$ are $l$-bit vectors, $X_1$ and $X_2$ are completions, and $i$ is a $m$-bit vector representing an input.
    \item The set of initial states of $M$ is $\{(s_0,X_1,s_0,X_2,i):X_1,X_2\in\mathcal{X},i\in\mathcal{I} \quad \textrm{and} \quad C_{X_1}(\mathbf{I})=C_{X_2}(\mathbf{I})=\mathcal{C}(\mathbf{I})\}$.
    \item $M$'s transition relation, $\mathcal{R}$ is defined as follows: $\mathcal{R}=((s_1,X_1,s_2,X_2,i),(s_1',X_1,s_2',X_2,i')): s_1,s_2,s_1',s_2'\in\mathcal{S}, \quad X_1,X_2\in\mathcal{X}, \quad i,i'\in\mathcal{I} \quad \textrm{such that} \quad \sigma_{X_1}(i,s_1)=s_1' \quad \textrm{and} \quad \sigma_{X_2}(i,s_2)=s_2'\}$. That is, the system can transition from a state $(s_1,X_1,s_2,X_2,i)$ to a state $(s_1',\allowbreak X_1,s_2',X_2,i')$ if and only if when $C_{X_1}$ and $C_{X_2}$ are in states $s_1$ and $s_2$, respectively, and we apply input $i$ to both circuits, $C_{X_1}$ and $C_{X_1}$ transition to states $s_1'$ and $s_2'$ respectively. The transition relation also requires that $X_1$ and $X_2$ retain their initial values throughout the evolution of the model.
    \item $M$'s labeling function, $\mathcal{L}$, is defined as follows: $L(s_1,X_1,\allowbreak s_2,X_2,i)=\{\texttt{equiv}\}$ if $\omega_{X_1}(s_1,i)=\omega_{X_2}(s_2,i)$, otherwise $L(s_1,X_1,s_2,X_2,i)=\emptyset$, i.e., the proposition \texttt{equiv} is true in a state $(s_1,X_1,s_2,X_2,i)$
    if and only if when $C_{X_1}$ and $C_{X_2}$ are in states $s_1$ and $s_2$, respectively, and we apply input $i$ to both circuits, $C_{X_1}$ and $C_{X_1}$ produce the same output.
    
\end{itemize}

We have the following proposition:

\begin{proposition}
For a given \discsetseqdec instance $\tuple{C,\mathbf{I},\mathcal{C}(\mathbf{I})}$, if the structure $M$ constructed as above satisfies the specification $\mathbf{G} \ \texttt{equiv}$, that is, for every $i\in\mathcal{I}$ and every $X_1,X_2\in\mathcal{X}$, $M,(s_0,X_1,s_0,X_2,i) \models \mathbf{G} \ \texttt{equiv}$, then $\tuple{C,\mathbf{I},\mathcal{C}(\mathbf{I})}$ is a true instance of \discsetseqdec, that is, $\mathbf{I}$ is a discriminating set for $C$.
\end{proposition}

\begin{proof}[(Sketch)]
We can interpret the structure $M$ as follows. $M$ represents the reachable states of a system that consists of two completed circuits $C_{X_1}$ and $C_{X_2}$ of $C$ that agree with the black-box circuit on every input sequence in $\mathbf{I}$. The two completions are simultaneously exercised at every step with the same input. The \texttt{equiv} proposition holds at any state if the two completions produce the same output for the input applied in that state. If the \texttt{equiv} proposition holds on every path of $M$, as expressed by the linear temporal logic formula $\mathbf{G} \ \texttt{equiv}$, then we have that any two such $C_{X_1}$ and $C_{X_2}$ are equivalent. Since a correct completion has to necessarily agree with the black-box circuit on every input sequence, it follows that both $X_1$ and $X_2$ are equivalent to \emph{some} correct completion, and are therefore, by extension, correct completions themselves. By the definition of a discriminating set of input sequences, then, it follows that $\tuple{C,\mathbf{I},\mathcal{C}(\mathbf{I})}$ is a true instance of \discsetseqdec.
\end{proof}

The above structure can be described using NuSMV's input language, with size at worst polynomial in the given \discsetseqdec instance. The $M_{UMC}$ solver thus produces a description of the structure $M$ corresponding to the input \discsetseqdec instance, and invokes a model checker asking to verify the specification $\mathbf{G} \ \texttt{equiv}$ on $M$. If the model checker says that $M$ satisfies the specification, $M_{UMC}$ returns \texttt{true}, otherwise $M_{UMC}$ parses the counterexample returned by the model checker for a tuple $\tuple{X_1,X_2,\tilde{I}}$ to be returned to the caller, as expressed in Algorithm~\ref{alg:reverse_eng}. We implement the solvers $M_{UC}$ and $M_{CE}$ using similar calls to a model checker on (slight variations of) the structure $M$.





\section{Experimental Results}
In this section, we describe our experimental results. 
We begin by describing our experimental setup, and then analyze the 
strength of our attack empirically. 

\subsection{Experimental Setup}

We implement our attack procedure using C++ in $\approx$ 700 lines of code, and use NuSMV~\cite{cimatti2000nusmv} 
as the back-end model checker. 
All experiments were executed on an Intel(R) Xeon CPU E5-2650 processor.  
We assume a camouflaged standard cell library 
that can implement either a NAND or a NOR function. Note that the camouflaged standard cell 
library in ~\cite{RSS+13} also implements XOR functions (in addition to NAND and NOR), but we did not observe any instances of
XOR in the benchmarks that we use.

\begin{table}[h]
\centering
\caption{Benchmark characteristics.}
\label{tab:bmarks}
\begin{tabular}{lllll}
\hline
B'mark & \#PIs  & \#POs & \#FFs & \#Gates \\ \hline
s344   & 9      & 11    & 15    & 160     \\
s349   &    9   & 11      & 15      & 161        \\
s382  & 3      &    6   & 21      & 158        \\ 
s400  & 3      &    6   & 21      & 162        \\ 
s444  & 3      &    6   & 21      & 181        \\ 
s526  & 3      &    6   & 21      & 193        \\ 
s820  & 18      & 19      & 5      &    289     \\ 
s832  & 18      & 19      & 5      &    287     \\ 
s953  & 16      & 23      & 29      & 395        \\ 
s1196  & 14       & 14      & 18      & 529        \\ 
s5378  & 35       & 49      & 179      &    2779     \\ 
s9234  &  19     & 22      &    228   & 5597        \\ 
s38584  &  12     & 278      & 1452      &  19253       \\ 
b04  & 11      &    8   & 66      & 628        \\ 
b08  & 9      & 4      &    21   & 183        \\ 
b14  & 32      &    54   & 245      & 5678        \\ \hline
\end{tabular}
\end{table}

We implemented two techniques to select which gates to 
camouflage: (1) the \emph{output corruptibility} + \emph{non-resolvable} (OC/NR) 
technique proposed by \cite{RSS+13},  
which is secure against VLSI test based attacks, and 
(2) random selection, in which the camouflaged gates are 
picked uniformly at random from the set of eligible gates in the circuit.

We use circuits from the ISCAS'89 \cite{iscas89} and the ITC'99 \cite{corno2000rt} sequential benchmark suites. 
The characteristics of the benchmarks in terms of the number of inputs, outputs, flip-flops and gates
is shown in Table~\ref{tab:bmarks}. More details about these benchmarks can be found in~\cite{iscas89} and \cite{corno2000rt}. 
We note that the b14 benchmark  implements a subset of the VIPER processor. For the s38584 benchmark, all our attack runs crashed, presumably because the model checking instances we generated in our procedure for s38584 were too large for NuSMV to handle. As such, we do not report any further results for this benchmark.

\subsection{Experimental Results}


Table~\ref{tab:rand} shows the results of our attack on the scheme in which gates are randomly camouflaged. 
We created $10$ different camouflaged circuits for every benchmark, each with a different random selection of $32$ 
camouflaged 
gates. 
The table plots (1) the number of discriminating input sequences, (2) the maximum length of an input sequence in the discriminating 
set, (3) the time taken by our attack, and (4) the number of attack runs (out of 10) that were successful and, 
for successful attacks, the termination condition that provided the correct completion. 

Several observations are in order. Out of 160 attacks, 135 runs were successful. We were unable to decamouflage any 
instance of the s9234 benchmark, only decamouflaged one instance of the s5378 benchmark, and seven instances of the s400 
and s444 benchmarks. 
On the other hand, our attack decamouflaged all instances of the b14 benchmark, one of the largest that we tried.

We hypothesize that even though the s400 and s444 benchmarks are small, they are hard 
to decamouflage because they have a relatively small number of primary inputs and outputs (3 PIs and 6 POs) compared to the 
number of bits of internal state (21 FFs each). We note that our attack required relatively long input sequences 
of length up to $90$ 
for the instances in which we successfully 
decamouflaged these benchmarks. 
We do show that on all of the instances of s400 and s444 for which we are unsuccessful, we are able to correctly 
recover at least 30 of 32 camouflaged gates. 

Another interesting observation is that our UC and CE
termination conditions that try to avoid calls to an unbounded model checker enhance
success of our attack. For example, on the one instance of c5378 that we decmaouflaged, we were able to terminate 
because our combinational equivalence (CE) check succeeded quickly; a call to unbounded model checker to
decide whether to terminate was still running after several hours.

Table~\ref{fig:ocnr} shows the same data, but this time for the OC/NR based camouflaging approach~\cite{RSS+13} and are qualitatively
similar. Since the procedure is deterministic, we generate only one camouflaged netlist for each benchmark. 
It is interesting to note that OC/NR does not seem to provide any additional security against the model checking 
attack compared to the random camouflaging scheme.


Our empirical attack results provide mixed evidence as to our central question: 
does blocking access to scan chains increase the immunity of camouflaged sequential circuits against 
reverse engineering attacks. On the one hand, using our proposed model checking attack, we were able to decamouflage 
relatively large benchmark circuits. On the other, benchmarks like s5378 and s9234
have been decamouflaged in 
prior work~\cite{el2015integrated} assuming scan access. Withholding scan access does seem to increase the security for these benchmarks.

\tabcolsep=0.11cm

\begin{table}[]
\centering
\caption{Results of proposed attack on FSMs camouflaged using random 
selection. Also noted are the termination conditions: unique completion (UC), 
combinational equivalence (CE) and unbounded model checker (UMC).}
\label{tab:rand}
\scriptsize
\begin{tabular}{|l|l|l|l|l|l|l|l|}
\hline
B'Mark & \multicolumn{2}{l|}{\# Disc Inputs} & \multicolumn{2}{l|}{Max Steps} & \multicolumn{2}{l|}{Time (s)} & Termination \\ \hline
       & min              & max              & min            & max           & min           & max           & UC/CE/UMC   \\ \hline
s344   & 3                  & 5                 &   10             & 10               & 11              & 37              & 10/0/0            \\ \hline
s349   &    3              &    7              & 10               & 10              & 15              & 69              & 10/0/0            \\ \hline
s382   &    25              &    36              & 50               & 60              & 3482              & 41129         & 10/0/0            \\ \hline
s400   &    18              & 34                 & 50               & 90              & 4921              & 526499         & 6//0/1            \\ \hline
s444   &    16              & 35                 &   50             & 90              & 3379              & 52984         & 2/0/5            \\ \hline
s510   &    7              &    15              & 30               & 40              & 300              & 29121         & 10/0/0            \\ \hline
s526   &    29              & 39                 & 120               & 120              &   37979            & 139252         & 10/0/0            \\ \hline
s820   &    14              & 20                 &  10              & 10              & 506              & 1030         & 10/0/0            \\ \hline
s832   &    12              & 21                 & 10               & 10              & 370              & 1211         & 10/0/0            \\ \hline
s953   &    10              & 22                 & 10               & 10              & 365              & 1709         & 10/0/0            \\ \hline
s1196   &   14               & 44                 & 10               & 10              &    795           &    2386      & 10/0/0            \\ \hline
s5378   &   7               & -                 & 30               & -              &    1350           &    -      & 0/1/0            \\ \hline
s9234   &   -               & -                 & -               & -              &    -           &    -      & 0/0/0            \\ \hline
b04   & 4                 & 9                 & 10               &    10           & 31              &    151      & 10/0/0            \\ \hline
b08   & 26                 &    117              &  20              &    20           & 619              & 10527         & 10/0/0            \\ \hline
b14   & 14                 &    21              & 10               &    10           & 14308              & 34273         & 10/0/0            \\ \hline
\end{tabular}
\end{table}

\begin{table}[]
\centering
\caption{Results of proposed attack on FSMs camouflaged using the OC/NR technique from \cite{RSS+13}. Also noted are the termination conditions: unique completion (UC), 
combinational equivalence (CE) and unbounded model checker (UMC).}
\label{fig:ocnr}
\scriptsize
\begin{tabular}{|l|l|l|l|l|l|l|l|}
\hline
B'Mark & \# Disc Inputs & Max Steps & Time (s)  & Termination   \\ \hline
s344   & 4              & 10        & 14        & UC            \\ \hline
s349   & 3              & 10        & 8               & UC            \\ \hline
s382   & 34 & 60 & 17713 & UMC \\ \hline
s400   & 26 & 60  & 14803 & UMC \\ \hline
s444   & 36 & 80 & 150569 & CE\\ \hline
s510   & 13 & 40 & 703 & UC \\ \hline
s526   & 29 & 80 & 13001 & UC \\ \hline
s820   & 23 & 10 & 1508 & UC \\ \hline
s832   & 16 & 10 & 253 & UC \\ \hline
s953   & 11 & 10 & 127 & UC \\ \hline
s1196   &   17               & 10                 & 1150 & UC            \\ \hline
s5378   &   -               & -                 & - & -            \\ \hline
s9234   &   -               & -                 & - & -            \\ \hline
b04   & 7 & 10 & 191 & UC \\ \hline
b08   & 31 & 20 & 1734 & UC \\ \hline
b14   &  13 &  40 &  19026 & UC \\ \hline
\end{tabular}
\end{table}

\subsubsection{Partial Completions}

\begin{figure}[h]
  \centering
  \includegraphics[width=0.8\columnwidth]{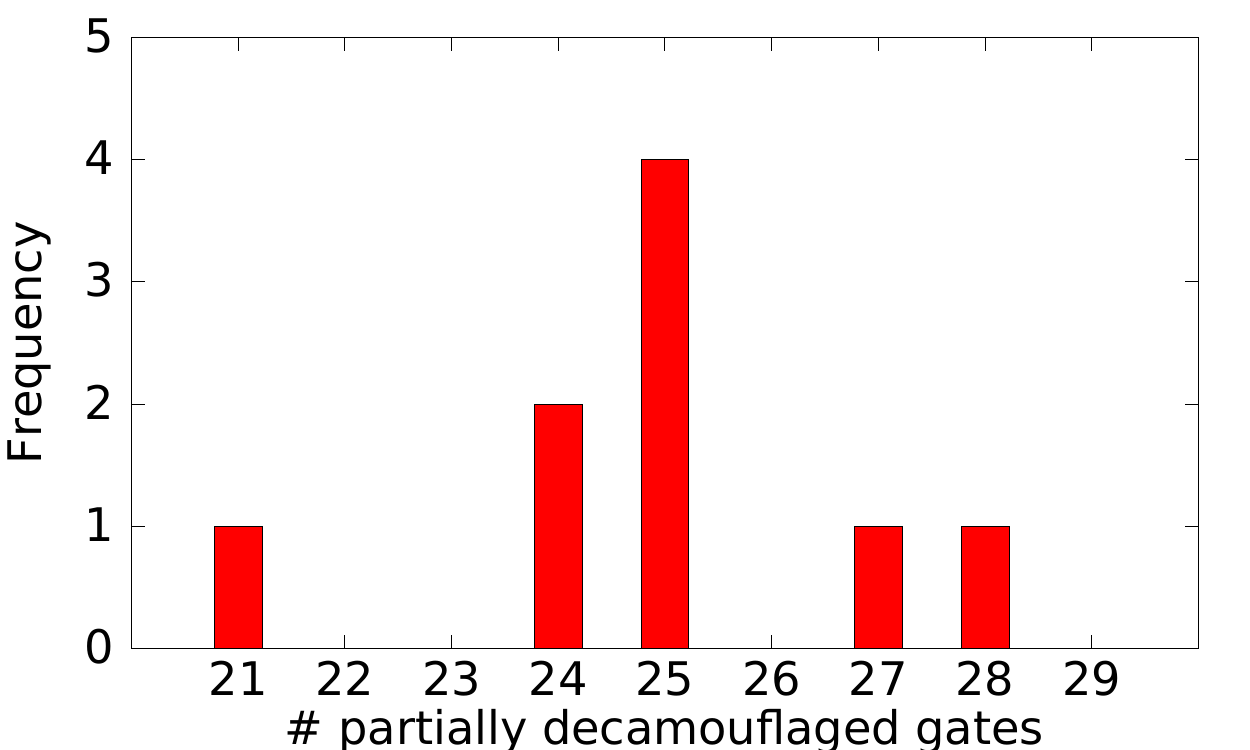}
  \caption{Histogram of number of partially decamouflaged gates for the s5378 benchmark across nine runs on which our attack 
  did not successfully recover every camouflaged gate.}
  \label{fig:hist}
\end{figure}

\begin{figure}[h]
    \centering
    \begin{subfigure}[b]{0.45\columnwidth}
      \includegraphics[width=\textwidth]{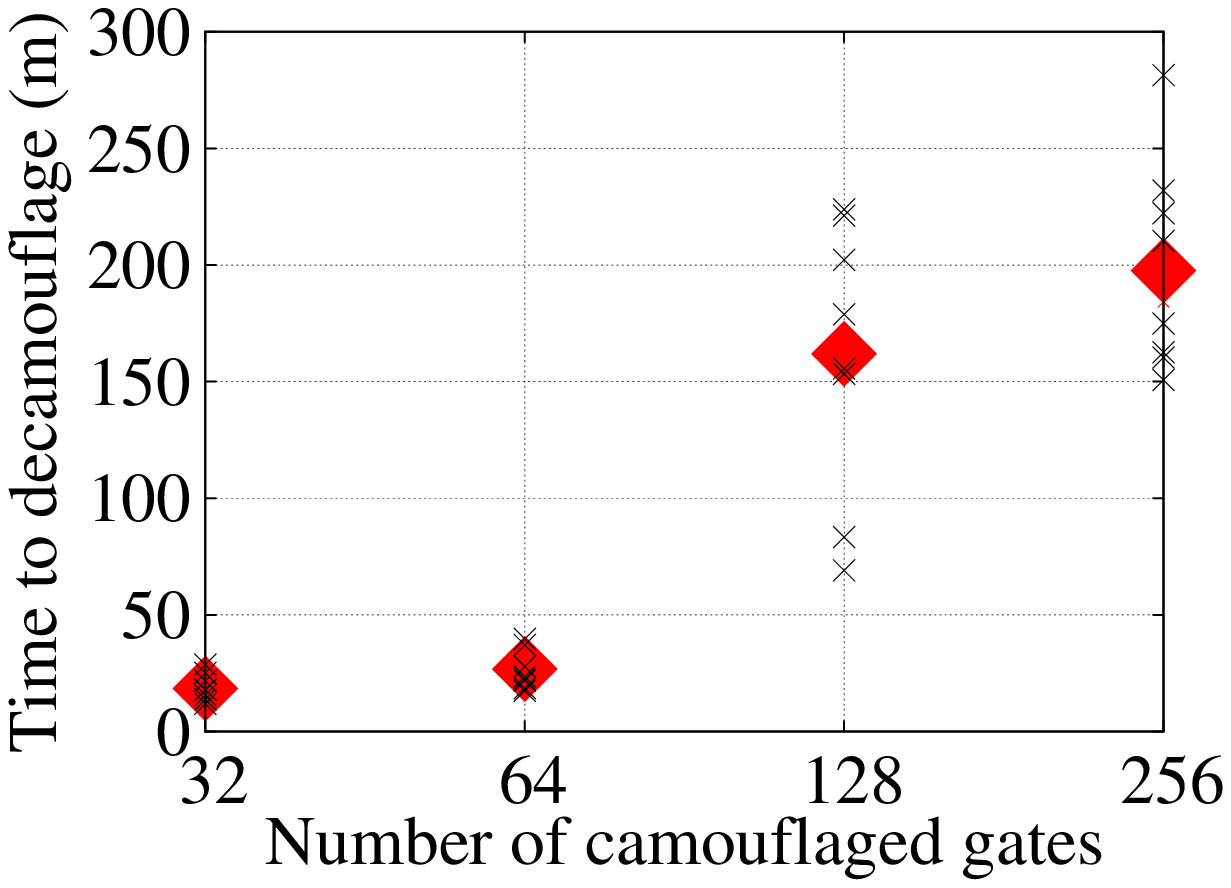}
      \caption{Attack time.}
      \label{fig:time_inc}
    \end{subfigure}
~
    \begin{subfigure}[b]{0.45\columnwidth}
      \includegraphics[width=\textwidth]{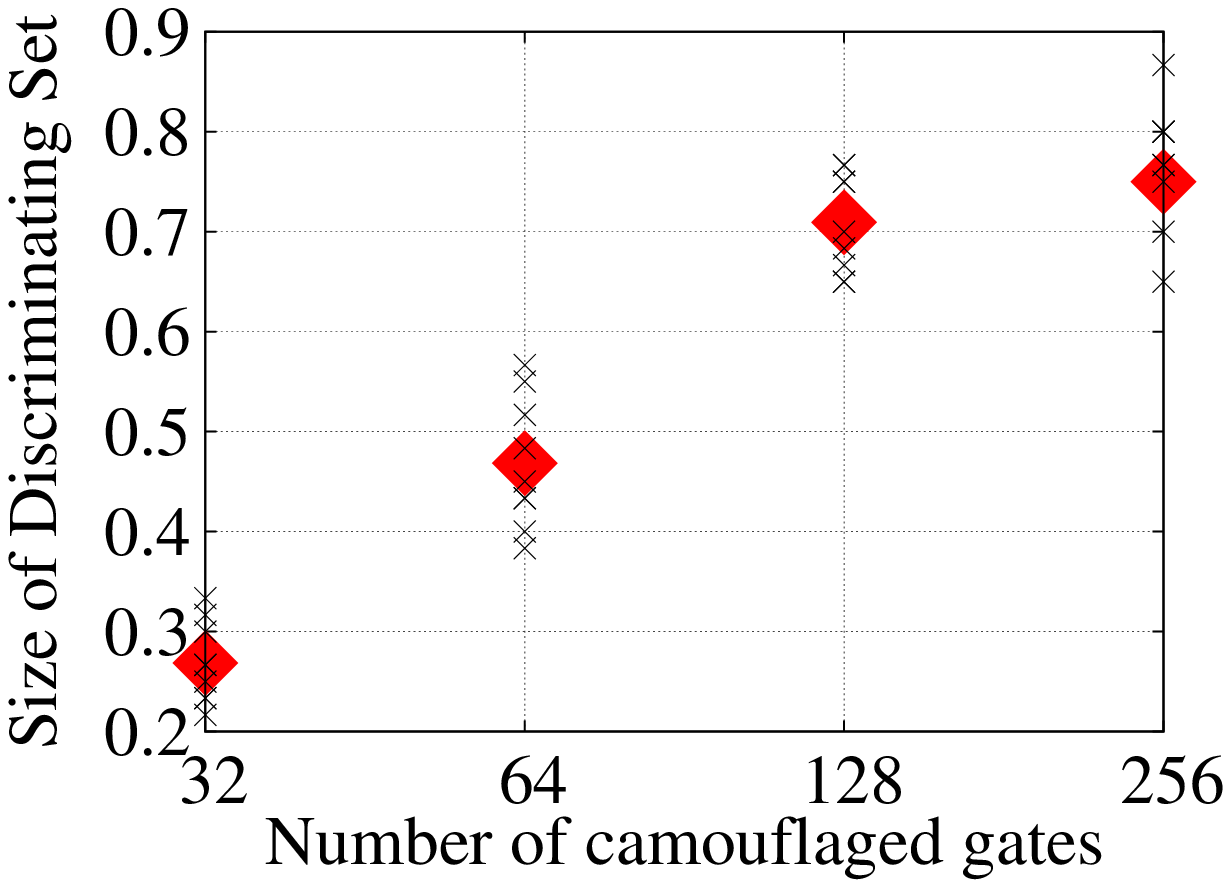}
      \caption{Discriminating set size.}
        \label{fig:disc_inc}
    \end{subfigure}
  \caption{Effect of increasing number of camouflaged gates on attack performance on the s1196 benchmark on (a) attack run-time, and 
  (b) size of the discriminating set of input sequences.}
  \label{fig:numcamo}
\end{figure}

We tried recovering the identities of as 
many camouflaged gates as possible for benchmarks that we could not successfully decamouflage every gate. 
We do this using a technique similar to the one proposed in \cite{subramanyan2015evaluating}, i.e., any camouflaged 
gate 
that is assigned the same identity by all remaining completions (those not eliminated by the set $\mathbf{I}$) 
can be assigned that identity. We call this a partial completion.

Based on this technique, we found that we were able to correctly identify 30 out of 32 camouflaged gates for the six
s400 and s444 benchmark instances that we could not fully decamouflage. 
Figure~\ref{fig:hist} plots the histogram of the number of correctly decamouflaged gates 
for the nine runs of the s5378 benchmark on which we were unsuccesful. We observe that between $21$ and $29$ out 
of $32$ gates are correctly decamouflaged, significantly reducing the attacker's search space. 
For instance, in the two cases 
where our partial completion attack recovered $29$ camouflaged gates, the attacker has only $16$ remaining possibilities. 

\subsubsection{Impact of Number of Camouflaged Gates}

So far, we have assumed in our experiments that only 32 gates are camouflaged.  We increased 
the number of camouflaged gates for the s1196 benchmark from 32 to 256 (including 10 runs for each) 
and found that we are able to decamouflage the circuit in each case. Figure~\ref{fig:numcamo} plots the run-time of our attack
and the number of input sequences required to decamouflage (size of discriminating set) each instance.  
An interesting observation is that although the number of input sequences required to decamouflage
the circuit increases with increasing number of camouflaged
gates, the length of the input sequences in the discriminating set was always at most $10$.

\section{Conclusion}

In this paper, we proposed the 
first attack methodology for reverse-engineering camouflaged sequential circuits 
without assuming that the attacker has scan chain access.  
We have identified the computational complexity of two underlying sub-problems 
on which our attack procedure relies, and show that the problem of determining when a 
given set of input sequences is sufficient to decamouflage a circuit is in \cpspace.
Based on this observation, we have developed a practical and scalable attack procedure that makes
iterative calls to a bounded model checker.  

Our attack is effective on the majority of the benchmarks we tested, including a large sequential benchmark with more than 5000 
gates. However, there are benchmarks that the attack does not fully decamouflage,
suggesting that removing scan access may 
indeed be helpful in increasing the resiliency of some circuits against 
reverse engineering attacks. The attack motivates the need for further research into 
camouflaging mechanisms for {sequential} circuits that leverage the attacker's lack of access to internal state to further  
enhance resilience against our attack.

\bibliographystyle{unsrt}
\bibliography{bibliography}

\appendix{Proof of Theorem \ref{thm:discsetdec}}

We prove that \discsetseqdec is in \cpspace by describing an algorithm for solving \discsetseqdec that requires an amount of space that is at worse polynomial in the size of the input instance. Recall that the input to \discsetseqdec is a tuple $C,\tuple{C,\mathbf{I},\mathbf{O}}$ where $C$ is a camouflaged circuit, $\mathbf{I}$ is a set of input sequences, and $\mathbf{O}$ is a set of output sequences corresponding to an input sequence in $\mathbf{I}$. A precondition on $C,\tuple{\mathbf{I}}$ is that there exists at least one completion $X$ of $C$ such that $C_X(\mathbf{I})=\mathbf{O}$. The algorithm we propose is as follows:

For every pair of completions $X_1$ and $X_2$ in $\mathcal{X}$, we check whether $C_{X_1}(\mathbf{I})=\mathbf{O}$ and $C_{X_2}(\mathbf{I})=\mathbf{O}$. This can be done using $O(|C|)$ space, where the size of $C$ is the number of inputs of wires plus the number of gates in $C$. If any of $X_1$ or $X_2$ do not satisfy the condition, we move on to the next pair; otherwise, we check whether $C_{X_1}$ and $C_{X_2}$ agree with each other on every input sequence of length $l$. Again, this can be done in $O(l|C|)$ space. If we find an input sequence of length $l$ for which $C_{X_1}$ and $C_{X_2}$ produce different outputs, we return \texttt{false}. If we exhaust every pair of completions in $\mathcal{X}$, we return \texttt{true}.

\end{document}